\documentclass[letterpaper, 10 pt, conference]{ieeeconf}  %

\IEEEoverridecommandlockouts                              %
\usepackage{lineno}
\usepackage{algorithm,algorithmic}
\usepackage{amssymb}
\usepackage{amsmath}
\usepackage{epsfig}
\usepackage{graphicx}
\usepackage{graphics}
\usepackage{float}
\usepackage{color}
\usepackage{lineno}
\usepackage{fullpage}
\usepackage[normalem]{ulem} 
\usepackage{makeidx}
\usepackage{xspace}
\usepackage{wrapfig}
\usepackage{mathtools}
\usepackage{cite}
\usepackage{hyperref}
\usepackage{xcolor}
\makeindex

\newtheorem{proposition}{Proposition}
\newtheorem{corollary}{Corollary}

\definecolor{darkgreen}{rgb}{0,0.5,0}

\usepackage[normalem]{ulem}

\usepackage{booktabs}
\usepackage{textgreek}
\usepackage{threeparttable}

\title{\LARGE \bf A Robust and Constrained Multi-Agent Reinforcement Learning Electric Vehicle Rebalancing Method in AMoD Systems} 
 
\author{
{Sihong He$^{1}$} \and {Yue Wang$^{2}$} \and {Shuo Han$^{3}$} \and {Shaofeng Zou$^{2}$} \and {Fei Miao$^{1}$}
\thanks{$^{1}$Sihong~He and Fei~Miao are with the Department of Computer Science and Engineering, University of Connecticut, Storrs, CT. {\tt\small \{sihong.he, fei.miao\}@uconn.edu}. }
\thanks{$^{2}$Yue Wang and Shaofeng Zou are with the Department of Electrical Engineering, University at Buffalo, The State University of New York. { \tt\small \{ywang294, szou3\}@buffalo.edu}. }
\thanks{$^{3}$Shuo Han is with the Department of Electrical and Computer Engineering, University of Illinois, Chicago. {\tt\small hanshuo@uic.edu}. }
\thanks{This work is supported by National Science Foundation under Grants CNS-1952096, CMMI-1932250, CNS-2047354, Grants CCF-2106560 and CCF-2007783.}
}
\begin{document}
\maketitle
\thispagestyle{plain}
\pagestyle{plain}

\begin{abstract}
Electric vehicles (EVs) play critical roles in autonomous mobility-on-demand (AMoD) systems, but their unique charging patterns increase the model uncertainties in AMoD systems (e.g. state transition probability). Since there usually exists a mismatch between the training and test/true environments, incorporating model uncertainty into system design is of critical importance in real-world applications.
However, model uncertainties have not been considered explicitly in EV AMoD system rebalancing by existing literature yet, and the coexistence of model uncertainties and constraints that the decision should satisfy makes the problem even more challenging. In this work, we design a robust and constrained multi-agent reinforcement learning (MARL) framework with state transition kernel uncertainty for EV AMoD systems. We then propose a robust and constrained MARL algorithm (ROCOMA) with robust natural policy gradients (RNPG) that trains a robust EV rebalancing policy to balance the supply-demand ratio and the charging utilization rate across the city under model uncertainty. Experiments show that the ROCOMA can learn an effective and robust rebalancing policy. It outperforms non-robust MARL methods in the presence of model uncertainties. It increases the system fairness by 19.6\% and decreases the rebalancing costs by 75.8\%.
\end{abstract}
\section{Introduction}
\label{sec_intro}
Autonomous mobility-on-demand (AMoD) system is one of the most promising energy-efficient transportation solutions as it provides people with one-way rides from their origins to destinations \cite{zardini2021analysis_survey}. Electric vehicles (EVs) are being adopted worldwide for environmental and economical benefits~\cite{IEA2020}, and AMoD systems embrace this trend without exception. However, the trips sporadically appear, and the origins and destinations are asymmetrically distributed. Such spatial-temporal nature of urban mobility motivates researchers to study vehicle rebalancing methods \cite{wen2017rebalancing,dro_he}, i.e. redistribution of vacant EVs to areas of high demand and assigning low-battery EVs to charging stations.

\begin{figure}
	\centering
	\includegraphics [width=0.93\linewidth]{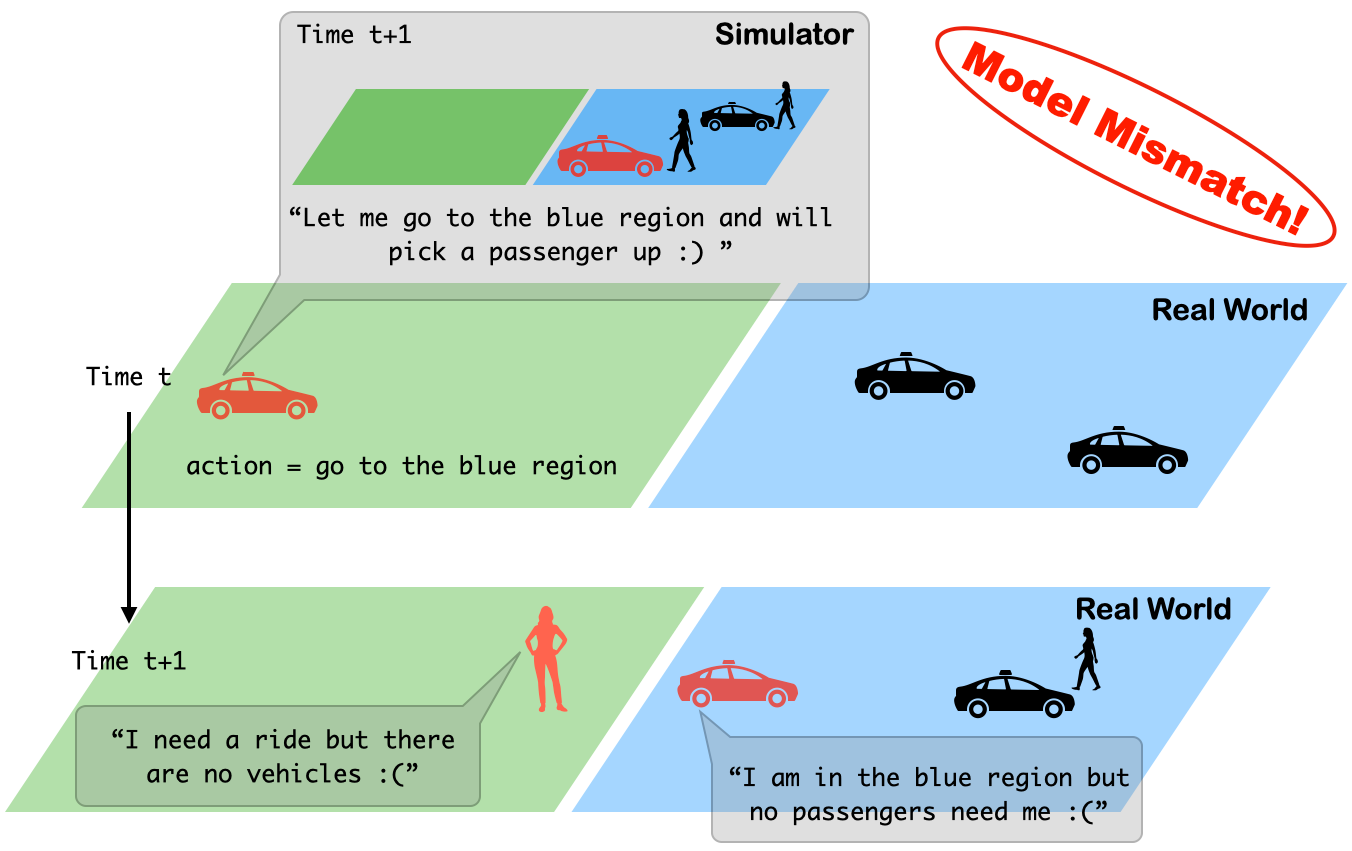}
	\vspace{-10pt}
	\caption{The model mismatch between the simulator and the real world degrades the performance of vehicle rebalancing methods. The red EV chooses to go to the blue region at time $t$ and thinks it can pick up a passenger at time $t+1$ according to the simulator model. However, in the real world, at time $t+1$, the red EV gets no passengers in the blue region and a passenger gets no cars in the green region.}
	\label{fig_mismatch}
 	\vspace{-25pt}
\end{figure}

In real-world AMoD systems, the simulation-to-reality gap remains challenging for vehicle rebalancing solutions calculated based on simulators, since there usually exists a model mismatch between the simulator (training environment) and the real world (test environment). For instance, at the current time, with the system state information such as the number of available vehicles and passenger demand in each region of the city, and the action to take as the number of available vehicles to be balanced among regions according to the mobility demand, it is difficult to accurately predict the state of the system (available vehicle supply and mobility demand) at the next time~\cite{zardini2021analysis_survey, dddro_tcps20, dro_Morari}. Hence, we usually do not have the true dynamic model of the system, i.e., the state transition probability. Thus, existing EV AMoD vehicle rebalancing methods~\cite{p2charge, EVcoodinate_19, pricing_20} may have significant performance degradation in the test (true) environment. One example is provided in Fig. \ref{fig_mismatch}.
Moreover, in real-world applications, the vehicle rebalancing decisions should satisfy specific constraints such as providing similar mobility and charging services in different regions which we call mobility and charging fairness. When there is a model mismatch, the rebalancing solution based on a simulator may violate the constraints in real AMoD systems.
Despite model-based methods considering prediction errors in mobility demand or vehicle supply ~\cite{mpcmod_icra16, dro_he,dddro_tcps20,hao2020robust_covariates}, how to calculate policies that satisfy the constraints and optimize the objectives %
under the uncertainty of state transition remains largely unexplored for AMoD rebalancing methods. 

In this work, to address the simulation-to-reality gap and calculate solutions that satisfy the fairness constraints, we propose a robust and constrained multi-agent reinforcement learning (MARL) framework for EV AMoD systems. The goal is to find robust policies that minimize the rebalancing distance of the vacant and low-battery EVs under model uncertainties and achieve mobility and charging fairness. The advantages of our methodology are two-fold: (i) fairness constraints can be satisfied even if there exist model uncertainty, and (ii) the expected rebalancing distance is optimized when there is a model mismatch. Our \textit{key contributions} are:

\textit{(1)} To the best of our knowledge, this work is the first to formulate EV AMoD system vehicle rebalancing as a robust and constrained MARL problem under model uncertainty. Via a proper design of the state, action, reward, cost constraints, and uncertainty set, 
we set our goal as minimizing the rebalancing distance while balancing the city's charging utilization and service quality, under model uncertainty.

\textit{(2)} %
We design a robust and constrained MARL algorithm (ROCOMA) to efficiently train robust policies. The proposed algorithm adopts the centralized training and decentralized execution (CTDE) framework. We also develop the robust natural policy gradient (RNPG) in robust and constrained MARL for the first time.

\textit{(3)} We run experiments based on real-world E-taxi system data. We show that our proposed algorithm performs better in terms of reward and fairness, which are increased by 19.6\%, and 75.8\%, respectively, compared with a non-robust MARL-based method when model uncertainty is present.

\section{Related Work}
\label{sec_related}
AMoD system vehicle rebalancing algorithms re-allocate vacant vehicles, sometimes considering charging constraints. Heuristics lead to sub-optimal rebalancing solutions \cite{liu2019dynamic_heuristic}. Other major categories of AMoD system rebalancing methods include optimization-based algorithms~\cite{he2023data}, Model Predictive Control (MPC) \cite{camacho2013model_mpc_book} and Reinforcement Learning (RL) \cite{sutton2018reinforcement_rl_book}. 

\textit{Optimization and MPC-based approaches} usually formulate the AMoD system vehicle rebalancing problem as an optimization problem, where the objective is to improve service quality~\cite{predictmod_icra17, Morari_rideshare} or maximize the number of served passengers with fewer vehicles~\cite{mpcmod_icra16, mod_iros18, DDmpcmod_icra18}. These model-based approaches usually rely on knowledge of the probability transition model of AMoD systems. Though robust and distributionally robust optimization-based methods have been designed to consider uncertainties caused by mobility demand, supply, or covariates predictions \cite{dro_he,hao2020robust_covariates}, the probability transition error or uncertainty in system dynamics has not been addressed yet. %
Various \textit{RL-based methods} include DQN, A2C and their variants \cite{wen2017rebalancing,holler2019deep_icdm,lin2018efficient,he2020spatio}
have been proposed to solve the vehicle rebalancing problem. However, RL suffers from the sim-to-real gap; that is, the gap between the simulator and the real world often leads to unsuccessful implementation if the learned policy is not robust to model uncertainties \cite{pinto2017robust}. None of the above RL-based rebalancing strategies consider this gap. %

\textit{Robust RL} has been proposed to find a policy that maximizes the worst-case cumulative reward over an uncertainty set of MDPs \cite{bagnell2001solving_uMDP, he2023robust, han2022solution}. 
To achieve a desired level of system fairness while minimizing rebalancing distance under model uncertainty, we put the fairness constraints in our RL formulation, which is known as \textit{Constrained RL} that aims to find a policy that maximizes an objective function while satisfying certain cost constraints \cite{wang2022policy_yue_iclr}. However, it remains challenging to design a robust EV rebalancing algorithm under model uncertainties and policy constraints, since the problem of robust constrained RL itself is already difficult to solve even in a simple tabular case. A robust and constrained RL for AMoD rebalancing cannot directly apply existing robust constrained RL solutions due to the high-dimensional state and action spaces commonly present in transportation systems. Our proposed robust and constrained MARL formulation and algorithm explicitly consider model uncertainties and policy constraints to learn robust rebalancing solutions for AMoD systems.

\section{Robust and Constrained MARL Framework for EV Rebalancing}
\label{sec_formulation}
\subsection{Preliminary: Multi-Agent Reinforcement Learning}
\label{subsec_marl}

We denote a Multi-Agent Reinforcement Learning (MARL) problem by a tuple $G = \langle \mathcal{N}, S, A, r,p,\gamma \rangle$, in which $\mathcal{N}$ is the set of $N$ agents. Each agent is associated with an action $a^i \in A^i$ and a state $s^i \in S$. We use $A = A^1 \times \cdots \times A^N$ to denote the joint action space, and $S = S^1 \times \cdots \times S^N$  the joint state space. At time $t$, each agent chooses an action $a^i_t$ according to a policy $\pi^i: S^i \rightarrow \Delta(A^i)$, where $\Delta(A^i)$ represents the set of probability distributions over the action set $A^i$. 
We use $\pi = \prod_{i=1}^N \pi^i: S \rightarrow \Delta(A)$ to denote the joint policy. After executing the joint action is executed, the next state follows the state transition probability which depends on the current state and the joint action, i.e.  $p: S \times A \rightarrow \Delta(S)$. And each agent receives a reward according to the reward function $r^i: S \times A \rightarrow \mathbb{R}$. Each agent aims to learn a policy $\pi^i$ to maximize its expected total discounted reward, i.e. $\max_{\pi^i} v^{\pi, i}_r(s)$ for all $s \in S$, where $v^{\pi, i}_r(s) = \mathbb{E}[\sum_{t = 1}^{\infty} \gamma^{t-1} r_t^i(s_t, a_t) | a_t \sim \pi(\cdot | s_t), s_1 = s]$ which is also known as the state value function for agent $i$. $\gamma \in (0,1)$ is the discounted rate. When these agents belong to a team, the objective of all agents is to collaboratively maximize the average expected total discounted reward over all agents, i.e. $\max_\pi v^{\pi}_{r}(s)$ for all $s \in S$, where $v^{\pi}_{r}(s) = \mathbb{E}_\pi[\sum_{t = 1}^{\infty} \gamma^{t-1} \sum_{i\in \mathcal{N}}r_t^i(s_t, a_t)/N | s_1 = s]$.

\subsection{Problem Statement}
We consider the problem of managing a large-scale EV fleet to provide fair and robust AMoD service. The goal is to (i) rebalance vacant EVs among different regions to provide fair mobility service on the passenger's side; (ii) allocate low-battery EVs to charging stations for fair charging service on the EVs' side; (iii) minimize the rebalancing distance of (i) and (ii). These three goals need to be achieved in the presence of model uncertainties, i.e. uncertainties in the state transition probability model of AMoD systems.

We divide the city into $N$ regions according to a pre-defined partition method~\cite{pricing_20, dro_he}. A day is divided into equal-length time intervals. In each time interval $[t, t+1)$, customers' ride requests and EVs' charging needs are aggregated in each region. After the location and status of each EV are observed, a local trip and charging assignment algorithm matches vacant EVs with passengers and low-battery EVs with charging stations, using existing methods in the literature~\cite{survey_19, carpool_cdc17}.
Then the state information of each region is updated, including the numbers of vacant EVs and available charging spots in each region. Each region then rebalances both vacant and low-battery EVs according to the well-trained MARL policy. This work focuses on a robust EV rebalancing algorithm design under model uncertainties to maximize the worst-case expected reward of the system while satisfying fairness constraints. 
For notational convenience, the parameters and variables defined in the following omit the time index $t$ when there is no confusion.

\subsection{Robust and Constrained Multi-Agent Reinforcement Learning Formulation for EV Rebalancing}
We formulate the EV rebalancing problem as a robust and constrained MARL problem $G_{rc} = \langle \mathcal{N}, S, A, P, r, c, d, \gamma \rangle$, and we define the agent, state, action, transition kernel uncertainty set, reward, and cost and fairness constraints as follows. %

\paragraph{\textbf{Agent}}The city is partitioned into a number of predetermined regions. Within each region, all the vacant and low-battery EVs are commanded by a single authority, which is referred to as a \textit{region agent}, whereas region agents from different regions independently make their rebalancing decisions of vacant and low-battery EVs. This multi-agent setting is more tractable for large-scale fleet management than a single-agent setting because the action space can be prohibitively large if we use a single system-wide agent~\cite{lin2018efficient}.

\paragraph{\textbf{State}}A state $s^{i}$ of a region agent $i$ consists two parts that indicate its spatiotemporal status from both the local view and global view of the city. We define the state $s^i=\{s_{loc}^i, s_{glo}^i\}$, where $s_{loc}^i = (V_i, L_i, D_i, E_i, C_i)$ is the state of region $i$ from the local view, denoting the number/amount of vacant EVs, low-battery EVs, mobility demand, empty charging spots, and total charging spots in region $i$, respectively. And $s_{glo}^i = (t, pos_i)$, where $t$ is the time index (which time interval), $pos_i$ is region location information (longitudes, latitudes, region index). The initial state distribution is $\rho$. 

\paragraph{\textbf{Action}}The rebalancing action for vacant EVs is denoted as $a^i_v  = \{ a^i_{v,j} \}_{j \in \text{Nebr}_i}$, the charging action for low-battery EVs as $a^i_l  = \{ a^i_{l,j} \}_{j \in \text{Nebr}_i}$, where $a^i_{v,j}, a^i_{l,j} \in [0,1]$ is the percentage of currently vacant EVs and low-battery EVs to be assigned to region $j$ from region $i$, respectively. And $\text{Nebr}_i$ is the set consisting of region $i$ and its adjacent regions as defined by the given partition. Therefore $\sum_{j \in \text{Nebr}_i} a^i_{v,j} = 1$ and $\sum_{j \in \text{Nebr}_i} a^i_{l,j} = 1$ for all $i$. We denote $m^i_{v,j} = h( a^i_{v,j} v^i)$ the actual number of vacant EVs assigned from region $i$ to region $j$, $m^i_{l,j} = h( a^i_{l,j} l^i)$ the actual number of low-battery EVs in region $i$ assigned to region $j$. The function $h(\cdot)$ is used to ensure that the numbers remain as integers and the constraints $\sum_j m^i_{v,j} = v^i, \sum_j m^i_{l,j} = l^i$ hold for all $i$.

\paragraph{\textbf{Transition Kernel Uncertainty Set}} 
We restrict the transition kernel $p$ to a $\delta$-contamination uncertainty set $P$ \cite{wang2022policy_yue_iclr}, in which the state transition could be arbitrarily perturbed by a small probability $\delta$. Specifically, let $\tilde{p} = \{ \tilde{p}^a_s \mid s \in S, a \in A \}$  be the centroid transition kernel, from which training samples are generated. The $\delta$-contamination uncertainty set centered at $\tilde{p}$ is defined as $P := \bigotimes_{s \in S, a \in A}P^a_s$, where $P^a_s := \{ (1-\delta)\tilde{p}^a_s + \delta q \mid q \in \Delta(S) \}, s \in S, a \in A$.

\paragraph{\textbf{Reward}}
Since one of our goals is to minimize the rebalancing distance, we define the shared reward as the negative value of the total rebalancing distance after EVs execute the decisions: $r(s,a) \coloneqq -[ d_v(s,a) + \bar\alpha d_l(s,a)]$, where $\bar\alpha$ is a positive coefficient, and $d_v(s,a), d_l(s,a)$ are moving distances of all vacant and low-battery EVs under the joint state $s$ and action $a$, respectively. We then define the worst-case value function of a joint policy $\pi$ as the worst-case expected total discounted reward under joint policy $\pi$ over $P$: $v^\pi_r(s) = \min_{p \in P} \mathbb{E}_\pi\left[ \sum_{t=1}^\infty \gamma^{t-1}{r}_t | s_1 = s \right]$. The notation is the same as MARL without considering uncertainty.
By maximizing the value function, region agents are cooperating for the same goal.

\paragraph{\textbf{Fairness Definition}}
We consider both the mobility supply-demand ratio~\cite{dddro_tcps20, Morari_rideshare} and the charging utilization rate~\cite{he2023data, EVcharge_19} in each region as service quality metrics. With limited supply volume in a city, keeping the supply-demand ratio of each region at a similar level allows passengers in the city to receive fair service~\cite{AMoD_queue, mpcmod_icra16}. Similarly, given a limited number of charging stations and spots, to improve the charging service quality and charging efficiency with limited infrastructures, balancing the charging utilization rate of all regions across the entire city is usually one objective in the scheduling of EV charging~\cite{EVcharge_19, EVcoodinate_19}.

The fairness metrics of the charging utilization rate $u_c$ and supply-demand ratio $u_m$ are designed based on 
the difference between the local and global quantities:
\begin{center}
$u_c(s,a) =   -\sum^N_{i = 1} \left\lvert  \frac{E_i}{C_i}- \frac{\sum^N_{j = 1}E_j}{\sum^N_{j = 1}C_j} \right\rvert$, 
$u_m(s,a)=  -\sum^N_{i = 1} \left\lvert  \frac{D_i}{V_i}- \frac{\sum^N_{j = 1}D_j}{\sum^N_{j = 1}V_j} \right\rvert$,
\end{center}
where $V_i$ is the number of vacant EVs in region $i$. The fairness metrics $u_s(s,a)$ and $u_m(s,a)$ are calculated given the EVs rebalancing action $a$, and the larger the better.  One advantage of the proposed robust and constrained MARL formulation is that the forms of the reward function and constraints do not need to satisfy the requirements as those of the robust optimization methods~\cite{he2023data, dddro_tcps20}, e.g., the objective/constraints do not need to be convex of the decision variable or concave of the uncertain parameters.

\paragraph{\textbf{Cost Function and Fairness Constraints}}
Another goal is to achieve the system-level benefit, i.e., balanced charging utilization and fair service. If the values of these fairness metrics are higher than some thresholds by applying a rebalancing policy $\pi$, we say the policy $\pi$ provides fair mobility and charging services among the city. We then augment the MARL problem $G$ with an auxiliary cost function $c$, and a limit $d$. The function $c: S \times A  \rightarrow \mathbb{R}$ maps transition tuples to cost, like the usual reward. Similarly, we let $v^{\pi}_c(s)$ denote the worst-case state value function of policy $\pi$ with respect to cost function $c$: $v^{\pi}_c(s) = \min_{p \in P}\mathbb{E}_\pi [\sum_{t=1}^{\infty} \gamma^{t-1} c(s_{t}, a_{t}) | s_1 = s]$. The cost function $c$ is defined as the system fairness (a weighted sum of city's charging fairness $u_c$ and mobility fairness $u_m$), i.e., $c(s,a) \coloneqq  u_c(s, a) + \bar\beta u_m(s, a)$, where $\bar\beta$ is a positive coefficient. Then the set of feasible joint policies for our robust and constrained MARL EV rebalancing problem is $\Pi_{C} := \{ \pi : \forall s\in S, v^{\pi}_c(s) \geq d \}$.

\paragraph{\textbf{Goal}}
The goal of our robust and constrained MARL EV rebalancing problem is to find an optimal joint policy $\pi^*$ that maximizes the worst-case expected value function subject to constraints on the worst-case expected cost:
\begin{align}
    \max_{\pi} \mathbb{E}_{s \sim \rho} [v^{\pi}_{r}(s)]
    \text{ s.t. }  \mathbb{E}_{s \sim \rho} [v^{\pi}_c(s)] \geq d
\end{align}
We define $v^{\pi_\theta}_{\text{tp}}(\rho)= \mathbb{E}_{s\sim\rho}[v^{\pi_\theta}_{\text{tp}}(s)]$, $\text{tp} \in \{r,c\}$. We then consider policies $\pi(\cdot | \theta)$ parameterized by $\theta$ and consider the following equivalent
max-min problem based on the Lagrangian \cite{Bookcvx_Boyd}:
\begin{align}
    \label{prob_roco_cmarl}
    \max_{\theta} \min_{\lambda \geq 0} J(\theta, \lambda) := v^{\pi_\theta}_r(\rho) + \lambda (v^{\pi_\theta}_c(\rho) - d),
\end{align}

\section{Algorithm}
\label{sec_algorithm}

\subsection{Robust and Constrained Multi-Agent Reinforcement Learning Algorithm (ROCOMA)}

\begin{figure}
    \centering
    \includegraphics[width=.51\textwidth]{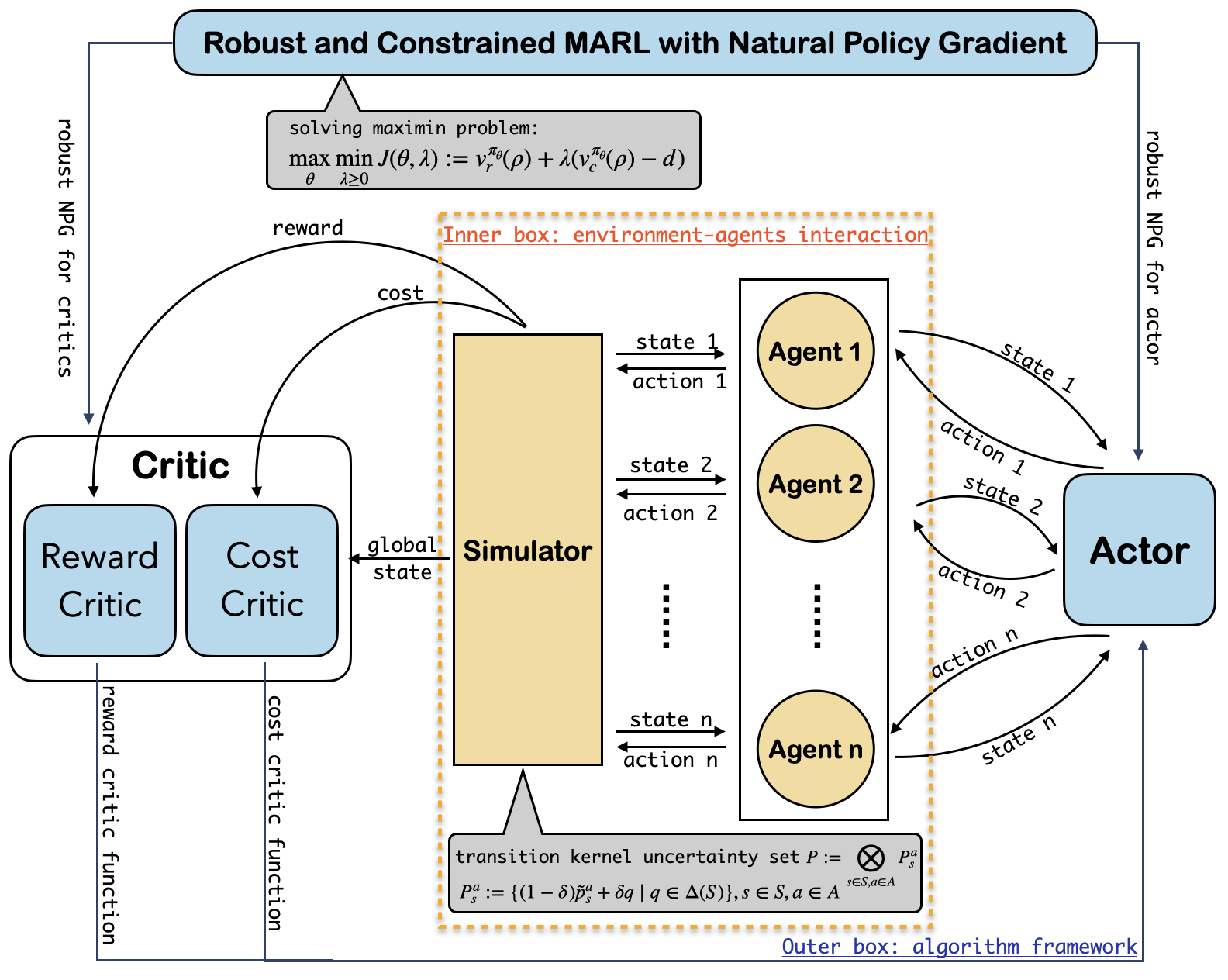}
    \vspace{-15pt}
    \caption{An Algorithm Overview for ROCOMA.}
    \label{fig_alg}
    \vspace{-15pt}
\end{figure}

We propose a robust and constrained MARL (ROCOMA) algorithm to solve the problem \eqref{prob_roco_cmarl} and train robust policies. The proposed algorithm is shown in Algorithm \ref{alg_rocoma_v4}. An algorithm overview is in Fig. \ref{fig_alg}. ROCOMA adopts the centralized training and decentralized execution (CTDE) framework, which enables us to train agents in the simulator using global information but executes well-trained policies in a decentralized manner in the real world. Specifically, we use centralized critic networks to approximate the value functions and decentralized actor networks to represent policies. Besides, for the first, time, we develop a robust natural policy gradient (RNPG) descent ascent to update actor networks and the Lagrange multiplier in MARL.  

As shown in Algorithm \ref{alg_rocoma_v4}, in line \ref{line_initialization}, we randomly initialize the actor network parameter $\theta_0$ and the Lagrange multiplier parameter $\lambda_0$. At each iteration $t$, in line \ref{line_estimate_value}, we estimate the critic networks $v_r^{\theta_t}, v_c^{\theta_t}$ under policy $\pi^{\theta_t}$ using Algorithm 3 in \cite{wang2022policy_yue_iclr}. Line \ref{alg_RNPG_start} to line \ref{alg_RNPG_end} are to estimate the robust natural policy gradient (RNPG) $\tilde{g}_{r,t}, \tilde{g}_{c,t}$ for $v_r^{\theta_t}$ and $v_c^{\theta_t}$, respectively. For notational convenience, we omit the subscripts $r$ and $c$ in the value functions when there is no confusion. In lines \ref{line_Time} and \ref{line_sample}, we sample an initial state $s^j_1$ following the initial distribution $\rho$ and a time horizon $T_j$ from the geometric distribution $\textit{Geom}(1-\gamma + \gamma\delta)$ at iteration $j=1, ..., M$. We use these samples to estimate the RNPG according to Corollary \ref{corollary_sgd}. Specifically, we initialize $\tilde{g}_{t,0}^j=0$ and use the following stochastic gradient descent (SGD) steps: $\tilde{g}_{t, k+1}^j = \tilde{g}_{t, k}^j-\zeta\nabla_{\tilde{g}} \mathcal{L}(\tilde{g}_{t,k}^j, \theta_t)$, where $\zeta$ is the learning rate and $\mathcal{L}(\tilde{g}_{t,k}^j,\theta_t) = \sum_{\mathcal{D}(s^j_{T_j})}[\tilde{g}^\top \psi^{\theta_t}(s,a) - \phi^{\theta_t}(\tau) - b^{\theta_t}]^2/D$, $\mathcal{D}(s^j_{T_j})$ is a set of trajectories $\tau$ starting at $s^j_{T_j}$ using policy $\pi^{\theta_t}$, i.e. $\tau = (s^j_{T_j},a,r,c,s^\prime)$, $D=|\mathcal{D}(s^j_{T_j})|$.
After $W$ steps of SGD iterations, the robust  natural policy gradient for $v^{\theta_t}(s^j_1)$ is estimated as $\sum_{k=1}^W\tilde{g}_{t,k}^j/W$.

To reduce the computational complexity, we adopt the centralized training and decentralized execution (CTDE) framework\cite{lowe2017multi} in ROCOMA and assume all agents share the same  policy $\pi^{\theta^i}(a^i|s^i)$, where $\theta^1=\cdots=\theta^N=\theta$. Then we have $\nabla \pi(a|s) = \sum_{i}^N \psi^{\theta}_i(s,a)$ where $\psi^{\theta}_i(s,a) := \pi^{-i}(a^{-i}|s^{-i})\nabla \pi^i(a^i|s^i)$, $\pi^{-i}(a^{-i}|s^{-i}):=\prod_{j\neq i}\pi^j(a^j|s^j)$. Therefore, in lines \ref{alg_ctde_start} to \ref{alg_ctde_end}, we address the high-dimensional action and state space issue in computing RNPG by using $\psi^{\theta}_i(s,a)$ instead of $\psi^{\theta}(s,a)$ in \eqref{RNPG_update}. 
Finally, we update $\theta_{t+1}$ and $\lambda_{t+1}$ using Gradient Descent Ascent (GDA)~\cite{lin2020gradient} in lines \ref{line_descent}, \ref{line_ascent}.

\subsection{Robust Natural Policy Gradient}
\label{subsec_rnpgda}

Natural policy gradient (NPG) \cite{schulman2015trust,lillicrap2015continuous} applies a preconditioning matrix to the gradient, and updates the policy along the steepest descent direction in the policy space\cite{ding2020natural_nips}. It has been proved that NPG moves toward choosing a greedy optimal action rather than just a better action \cite{kakade2001natural_nips}.  Generally, for a function $L$ defined on a Riemannian manifold $\Theta$  with a metric $M$, the steepest descent direction of $L$ at $\theta$ is given by $-M^{-1}(\theta)\nabla L(\theta)$, which is called the natural gradient of $L$ \cite{amari1998natural}. In the policy parameter space $\left\{ \pi_\theta \right\}$, the natural gradient of $L$ at $\theta$ is given by $\tilde{\nabla} L(\theta)=F(\theta)^{-1}\nabla L(\theta)$, where $F(\theta) := \mathbb{E}_{s} \left[ F_s(\theta) \right]$ is the Fisher information matrix at $\theta$ and $F_s(\theta) = \mathbb{E}_{\pi(a|s, \theta)} \left[ \frac{\partial \log \pi(a|s, \theta)}{\partial \theta_i} \frac{\partial \log \pi(a|s, \theta)}{\partial \theta_j} \right]$ \cite{kakade2001natural_nips}. Although the natural gradient method has been studied in non-robust RL, it is not straightforward to efficiently find the NPG for a robust and constrained MARL problem. %
We show the robust natural policy gradient for robust and constrained MARL in the following Proposition \ref{theorem_rnpg}.

\begin{algorithm}
\caption{Robust and Constrained Multi-Agent Reinforcement Learning Algorithm (ROCOMA)}
\label{alg_rocoma_v4}
    \begin{algorithmic}[1]
        \label{line_initialization}
        \STATE Input $\zeta,\alpha,\beta,\gamma,\delta$. Initialize $\theta_0, \lambda_0$.
        \FOR{$t=0$ to $T$}
            \STATE Estimate $v_r^{\theta_t}, v_c^{\theta_t}$ using Algorithm 3 in \cite{wang2022policy_yue_iclr}
            \label{line_estimate_value}
            \FOR{$j=1$ to $M$} \label{alg_RNPG_start}
                \STATE Sample $T_j \sim \textit{Geom}(1-\gamma+\gamma\delta)$, $s_1^j \sim \rho$
                \label{line_Time}
                \STATE Sample trajectory from $s^j_1$: $(s^j_1, a^j_1, \cdots, s^j_{T_j})$
                \label{line_sample}
                    \FOR{agent $i=1$ to $N$} \label{alg_ctde_start}
                    \FOR{$k=1$ to $W$}
                        \STATE $\tilde{g}_{t, k+1}^j(i) = \tilde{g}_{t, k}^j(i) -\zeta\nabla_{\tilde{g}} \mathcal{L}(\tilde{g}_{t,k}^j(i), \theta_t)$, $\mathcal{L}$ is defined in \eqref{RNPG_update}
                    \ENDFOR
                    \STATE $\tilde{g}^j_{t,k} = \sum_{i=1}^N \tilde{g}^j_{t,k}(i)/N$
                \ENDFOR \label{alg_ctde_end}
            \ENDFOR
            \STATE $\tilde{g}_t = \sum_{j=1}^M\sum_{k=1}^W \tilde{g}^j_{t,k}/MW$ \label{alg_RNPG_end}
            \STATE $\theta_{t+1} = \theta_t + \alpha_t(\tilde{g}_{r,t} + \lambda_t \tilde{g}_{c,t})$
            \label{line_descent}
            \STATE $\lambda_{t+1} = \max\{\lambda_t - \beta_t( \sum_{j}v_c^{\theta_t}(s_1^j)/M-d),0\}$
            \label{line_ascent}
        \ENDFOR
        \STATE Output $\theta_T$
    \end{algorithmic}
\end{algorithm}

\begin{proposition}[Robust Natural Policy Gradient]
\label{theorem_rnpg}
\textit{Let $\tilde{g}^*$ minimizes the objective $J(\tilde{g},\pi_\theta)$ defined as follows:
\begin{align}
    \sum_{s,a}d^\pi_{\gamma,\delta,s_1}\pi(a|s)
    [\tilde{g}^\top \psi^\pi(s,a) - \phi^\pi(\tau) - b^\pi]^2,
    \label{def_se}
\end{align}
where $d^\pi_{\gamma, \delta, s_1}$ $\propto \sum_k\gamma^k(1-\delta)^k p^\pi(s_k=s|s_1)$ is the discounted visitation distribution of $s_{k}=s$ when the initial state is $s_1$ and policy $\pi$ is used; $\psi^\pi(s,a)$ denotes $\nabla \log \pi(a|s,\theta)$; $\tau$ denotes a trajectory $(s,a,r,c,s^\prime)$; $\phi^\pi(\tau) = r+ \gamma\delta\min_sv^\pi(s)+\gamma(1-\delta)v^\pi(s^\prime)-v^\pi(s)$ is the TD residual; $b^\pi ={\gamma\delta}/{(1-\gamma+\gamma\delta)}\partial_\theta\min_{s}v^\pi(s)$.}

\textit{Then $\tilde{g}^*= F(\theta)^{-1}\nabla_\theta v^\pi(s_1)$ is the robust natural policy gradient of the objective function $v^\pi(s_1)$. For notational
convenience, we omit the subscripts $r$ and $c$ in the value functions when there is no confusion.}
\end{proposition}

\begin{proof}
Considering we have denoted $\psi^\pi(s,a) = \nabla \log \pi(a|s,\theta)$, Fisher information matrix is then given by $F(\theta) = \sum_{s,a}d^\pi_{\gamma,\delta,s_1}(s)\pi(a|s)\psi^\pi(s,a)\psi^\pi(s,a)^\top$.
The robust policy gradient of the value function is given by $\nabla_\theta v^\pi(s_1) = \sum_{s,a} d^\pi_{\gamma,\delta,s_1}(s) \nabla_\theta \pi(a|s)\phi^\pi(\tau) + b^\pi\propto \mathbb{E}_{\pi,s_1}[\phi^\pi(\tau)\nabla \log \pi(a|s)+b^\pi]$ \cite{wang2022policy_yue_iclr}.

Since $\tilde{g}^*$ minimizes \eqref{def_se}, it satisfies the condition $\partial J/\partial \tilde{g}_i = 0$, which implies: $\sum_{s,a}d^\pi_{\gamma,\delta,s_1}\pi(a|s) \times\psi^\pi(s,a)[ \psi^\pi(s,a)^\top \tilde{g}^* - \phi^\pi(\tau) -b^\pi] = 0.$
Then we have
\begin{align}
    &\sum_{s,a}d^\pi_{\gamma,\delta,s_1}\pi(a|s) \psi^\pi(s,a)\psi^\pi(s,a)^\top \tilde{g}^*
     \\
    = &\sum_{s,a}d^\pi_{\gamma,\delta,s_1}\pi(a|s) \psi^\pi(s,a)[ \phi^\pi(\tau) +b^\pi].\nonumber
\end{align}
By the definition of Fisher information:
$\text{LHS} = F(\theta)\tilde{g}^*$ and $\text{RHS} = \nabla_\theta v^\pi(s_1)$, which lead to: $F(\theta) \tilde{g}^* = \nabla_\theta v^\pi(s_1)$.
Solving for $\tilde{g}^*$ gives $\tilde{g}^* = F(\theta)^{-1}\nabla_\theta v^\pi(s_1)$ which follows from the definition of the NPG on the worst-case value function of robust and constrained MARL. We name it a robust natural policy gradient in robust and constrained MARL.
\end{proof}

Considering the vanilla policy gradient may suffer from overshooting or undershooting and high variance, which results in slow convergence \cite{liu2020improved}, our proposed robust natural policy gradient (RNPG) method updates the policy along the steepest ascent direction in the policy space in robust and constrained MARL \cite{ding2020natural_nips}. In Corollary \ref{corollary_sgd}, we show how to efficiently calculate RNPG by stochastic gradient descent (SGD).

\begin{corollary}[Calculating RNPG by SGD]
\label{corollary_sgd}
\textit{As shown in Proposition \ref{theorem_rnpg}, we can get the RNPG of $v^\pi(s_1)$ by minimizing the objective defined in \eqref{def_se}. To minimize \eqref{def_se} and get the minimizer, we initialize $\tilde{g}_{0}=0$ and use the following stochastic gradient descent (SGD) steps: 
$$\tilde{g}_{k+1} = \tilde{g}_{k}-\zeta\nabla_{\tilde{g}} \mathcal{L}(\tilde{g}_{k}, \pi),$$
where $\zeta$ is the learning rate and $\mathcal{L}$ is defined as follows:
\begin{align}
    \label{RNPG_update}
    \mathcal{L}(\tilde{g},\pi) = \sum_{\mathcal{D}(s_1)}[\tilde{g}^\top \psi^{\pi}(s,a) - \phi^\pi(\tau) - b^\pi]^2/D,
\end{align}
where $\mathcal{D}(s_1)$ is a set of trajectories $\tau$ starting at $s_1$ using policy $\pi$, i.e. $(s_1,a,r,c,s^\prime)$, $D=|\mathcal{D}(s_1)|$.
After $W$ steps of SGD iterations, the robust  natural policy gradient for $v^{\pi}(s_1)$ is estimated as $\sum_{k=1}^W\tilde{g}_{k}/W$.}
\end{corollary}

\color{red}

\color{black}

\section{Experiment}
\label{sec_experiment}

\subsection{Experiment Setup} 
Three different data sets~\cite{dro_he} including E-taxi GPS data, transaction data, and charging station data are used to build an EV AMoD system simulator as the training and testing environment. We infer parameters of order/charging station/vehicle generation models \cite{liu2020improved} from these data sets, then generate (1) the locations and availability of charging stations, (2) the number, origins, and destinations of mobility demand, (3) the initial location and state of charging of EVs. We modify the parameters of the simulator model such that the testing environment is different from the training environment, e.g., the parameters of the order generator. The simulated map is set as a grid city. The policy networks and critic networks are two-layer fully-connected networks, both with 32 nodes. We use Softplus as activations to ensure the output is positive. The output of policy networks is used to be the concentration parameters of the Dirichlet distribution to satisfy the action constraints (sum to one). We set the maximal training episode number $=20000$, the maximal policy/critic estimation number $=2000$, the RNPG SDG iteration number $=500$, the discount rate $\gamma= 0.99$, the perturbed rate $\delta=0.05$, the coefficients $\bar\alpha=\bar\beta=1$, the fairness constraint limit $d=-20$ for one simulation step, and use Adam optimizer with a learning rate of $0.001$ for both policy and critic networks.

\subsection{Experiment Results}
Our goal of the experiments is to validate the following hypothesis: (1) The proposed ROCOMA can learn effective rebalancing policies; (2) Our proposed ROCOMA learns more robust policies than a non-robust MARL algorithm does by considering state transition uncertainties and constraints in the MARL problem formulation and the proposed RNPG method for policy training. We compare metrics: \textit{Rebalancing distance:} the total moving distance of vacant and low-battery EVs by using a rebalancing policy (the lower the better); and \textit{System fairness:} the weighted sum of mobility and charging fairness (the higher the better); we also monitor \textit{Number of expired orders:} the total number of canceled orders due to waiting for more than $20$ minutes (the lower the better) and \textit{Order response rate:} the ratio between the number of served demands and the number of total passenger demand (the higher the better). 
All metrics are calculated in every testing period which consists of $25$ simulation steps. 
Then the fairness constraint limit for one testing period is $-500$.
We repeat testing for $10$ times and show the average values. 
\begin{table}[]
\centering
\begin{threeparttable}
\caption{Comparison: ROCOMA VS other rebalancing methods}
\vspace*{-10pt}
\label{tab_other}
\begin{tabular}{c|cccc}
\hline
  & rebalancing  & system & expired  & response  \\
  &  cost & fairness &  order &  rate \\ \hline
ROCOMA        & $2.06\times10^5$           & $-292.14$  & $1.20\times10^2$            & $99.82\%$         \\
MADDPG        & $1.94\times10^5$           & $-679.72$  & $3.46\times10^3$            & $85.06\%$         \\
COP        & $1.88\times10^5$           & $-383.19$  & $1.61\times10^3$          & $93.05\%$         \\
EDP        & $2.15\times10^5$           & $-409.49$  & $6.90\times10^1$            & $99.69\%$         \\
RDP        & $2.43\times10^5$           & $-629.85$  & $3.68\times10^3$          & $84.34\%$         \\
NO  & -                  & $-4317.53$ & $7.64\times10^3$          & $66.89\%$  \\
\hline
\end{tabular}
\begin{tablenotes}
\item[1]Compared to no rebalancing, by using our method, the expired orders number is decreased by 98.4\%, the system fairness and order response rate are increased by about 93.2\% and 32.9\%, respectively.
\end{tablenotes}
\vspace*{-15pt}
\end{threeparttable}
\end{table}

\paragraph{ROCOMA is effective} In Table \ref{tab_other}, we compare ROCOMA with \textit{no rebalancing scenario (NO)}, \textit{multi-agent deep deterministic policy gradient (MADDPG)} which is a state-of-the-art non-robust MARL algorithm \cite{lowe2017multi}, and  the following rebalancing algorithms: 
\textit{(1) Constrained optimization policy (COP):} The optimization goal is to minimize the rebalancing distance under the fairness constraints \cite{dro_he}. The fairness limit is the same as that used in ROCOMA. The dynamic models are calculated from the same data sets used in simulator construction.
\textit{(2) Equally distributed policy (EDP):} EVs are assigned to their current and adjacent regions using equal probability (20\%).
\textit{(3) Randomly distributed policy (RDP):} EVs are randomly distributed to their current and adjacent regions.

\begin{figure}
    \centering
    \includegraphics[width=.46\textwidth]{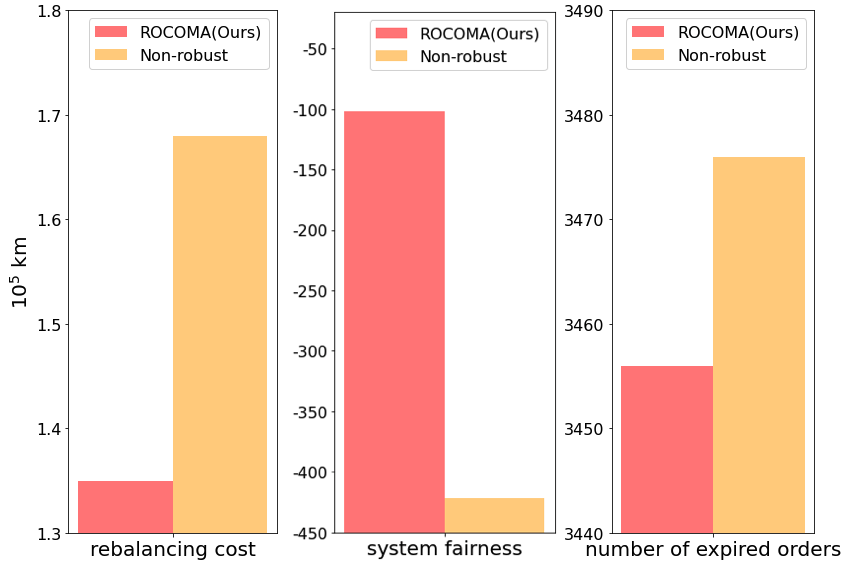}
    \vspace{-15pt}
    \caption{Comparison of ROCOMA and Non-robust MARL method: Compared to the non-robust method, ROCOMA decreases the rebalancing distance and increases the system fairness by 19.6\% and 75.8\%, respectively, when model uncertainties are present.}
    \label{fig_robust}
    \vspace{-15pt}
\end{figure}

In Table \ref{tab_other}, compared to the no rebalancing scenario, ROCOMA is effective in rebalancing AMoD systems in terms of fairness, expired orders and response rate. Specifically, ROCOMA policy decreases the number of expired orders by $98.4\%$, and increases the system fairness and order response rate by about $93.2\%$ and $32.9\%$, respectively. Besides, ROCOMA achieves a higher system fairness and order response rate using less rebalancing distance than EDP and RDP. Though ROCOMA takes more rebalancing distances than COP and MADDPG, it has a better system fairness and order response rate. It is within expectation since the constrained optimization method is a centralized method that aims to optimize the rebalancing distance and it does not consider any uncertainties.

\paragraph{ROCOMA is robust} In Figure \ref{fig_robust} and Table \ref{tab_rl}, we compare ROCOMA with \textit{(1) Non-constrained MARL algorithm:} Instead of considering fairness constraints in MARL, the reward is designed as a weighted sum of negative rebalancing distance and system fairness. The coefficient is $1$. And model uncertainty is considered; \textit{(2) Non-robust MARL algorithm:} The model uncertainty is not considered but the fairness constraint is considered in MARL. They use the same network structures and other hyper-parameters as that in ROCOMA.

In Figure \ref{fig_robust}, we test well-trained robust and non-robust methods in a testing environment (different from the training environment) to show the robustness of the ROCOMA policy. We can see ROCOMA policy achieves better performance in terms of all metrics. Specifically, ROCOMA decreases the rebalancing distance and increases the system fairness by about 19.6\% and 75.8\% , respectively, when model uncertainty exists, compared to the non-robust method. 

\begin{table}[]
\centering
\begin{threeparttable}
\caption{Comparison: ROCOMA VS non-constrained MARL method}
\label{tab_rl}
\begin{tabular}{c|cccc}
\hline
  & rebalancing  & system & expired  & response  \\
  &  cost & fairness &  order &  rate \\ \hline
ROCOMA        & $2.06\times10^5$           & $-292.14$  & $120$            & $99.82\%$\\
Non-constrained        & $1.98\times10^5$           &  $-1812.48$      &  $1607$         & $93.06\%$         \\
\hline
\end{tabular}
\begin{tablenotes}
\item[3] Our method achieves 83.9\% higher in fairness compared to the non-constrained MARL method with 4\% extra rebalancing distance.
\vspace*{-15pt}
\end{tablenotes}
\end{threeparttable}
\end{table}

In Table \ref{tab_rl}, ROCOMA achieves $83.9\%$ higher in fairness compared to the non-constrained MARL algorithm with just $4\%$ extra rebalancing distance. Without the fairness constraint design, the non-constrained MARL method falls into a pit that sacrifices fairness to achieve a lower rebalancing distance since its objective is a weighted sum of them. It would take a lot of effort to tune the hyper-parameter to find a policy that performs well in both rebalancing distance and fairness. The constrained MARL design of ROCOMA avoids such extra tuning efforts.

\section{Conclusion}
\label{sec_conclusion}
It remains challenging to address AMoD system model uncertainties caused by EVs' unique charging patterns and AMoD systems' mobility dynamics in algorithm design. In this work, we design a robust and constrained multi-agent reinforcement learning framework to balance the mobility supply-demand ratio and the charging utilization rate, and minimize the rebalancing distance for EV AMoD systems under state transition uncertainties.  We then design a robust and constrained MARL algorithm (ROCOMA) to train robust policies. Experiments show that our proposed robust algorithm can learn effective and robust rebalancing policies.

{ 
\bibliographystyle{IEEEtran}
\bibliography{008_sec_ref}
}

\end{document}